\documentclass{article}

\usepackage{arxiv}

\newlength\savedwidth

\usepackage{multirow}
\usepackage{array}
\usepackage[ruled,vlined]{algorithm2e}

\usepackage{xcolor}

\usepackage[utf8]{inputenc} 
\usepackage[T1]{fontenc}    
\usepackage{hyperref} 
\usepackage{scalerel,stackengine}
\usepackage{amsmath}
\stackMath
\newcommand\reallywidehat[1]{%
\savestack{\tmpbox}{\stretchto{%
  \scaleto{%
    \scalerel*[\widthof{\ensuremath{#1}}]{\kern-.6pt\bigwedge\kern-.6pt}%
    {\rule[-\textheight/2]{1ex}{\textheight}}
  }{\textheight}%
}{0.5ex}}%
\stackon[1pt]{#1}{\tmpbox}%
}

\usepackage{graphicx}
\usepackage[export]{adjustbox}
\usepackage{xcolor}
\usepackage{subfigure}
\usepackage{colortbl}
\usepackage{paralist}
\usepackage{bbm}
\usepackage{url}            
\usepackage{booktabs}       
\usepackage{amsfonts}       
\usepackage{nicefrac}       
\usepackage{microtype}      
\usepackage{lipsum}	
\usepackage{xcolor}
\title{Exploring Polarization of Users Behavior on Twitter During the 2019 South American Protests}

\author{
Ramon Villa-Cox \\
  \small{Carnegie Mellon University} \\
  \texttt{rvillaco@andrew.cmu.edu} \\
  \And
  Helen (Shuxuan) Zeng\\
  \small{Carnegie Mellon University}\\
  \texttt{shuxuanz@andrew.cmu.edu} \\
\And
Ashiqur R. KhudaBukhsh \\
  \small{Carnegie Mellon University}\\
  \texttt{akhudabu@cs.cmu.edu} \\
 \And
Kathleen M. Carley \\
  \small{Carnegie Mellon University}\\
  \texttt{kathleen.carley@cs.cmu.edu} \\
}

\begin{document}
\maketitle

\begin{abstract}

Research across different disciplines has documented the expanding polarization in social media. However, much of it focused on the US political system or its culturally controversial topics. In this work, we explore polarization on Twitter in a different context, namely the protest that paralyzed several countries in the South American region in 2019. By leveraging users’ endorsement of politicians' tweets and hashtag campaigns with defined stances towards the protest (for or against), we construct a weakly labeled stance dataset with millions of users. We explore polarization in two related dimensions: language and news consumption patterns. In terms of linguistic polarization, we apply recent insights that leveraged machine translation methods, showing that the two communities speak consistently ``different'' languages, mainly along ideological lines (e.g., fascist translates to communist). Our results indicate that this recently-proposed methodology is also informative in different languages and contexts than originally applied. In terms of news consumption patterns, we cluster news agencies based on homogeneity of their user bases and quantify the observed polarization in its consumption. We find empirical evidence of the “filter bubble” phenomenon during the event, as we not only show that the user bases are homogeneous in terms of stance, but the probability that a user transitions from media of different clusters is low.

\end{abstract}

\keywords{Linguistic Polarization \and Filter Bubbles \and 2019 South American Protests}

\section{Introduction}
The 2019 South American Protest were a series of protests that shocked the region at the end of 2019. These protests started in Ecuador and were followed by Chile, Bolivia and Colombia and effectively paralyzed the countries for months. With the exception of Bolivia, the protest resulted from populist movements seeking to resist austerity measures being imposed in each country and demanding more government spending in social programs. In Bolivia, they were response to an alleged electoral fraud undertaken by the government in favor of the president who was seeking reelection. These protests also had in common a massive online presence and the reported involvement of international and regional actors that sought to influence their evolution. These include international news agencies like RT en Español, funded in part by the Russian government,  TeleSUR and NTN24, funded in part by the Venezuelan government, that were more critical of local governments (except for Bolivia) and provided more favorable coverage of the protesters. In contrast, local news agencies tended to be more critical of them and favorable towards the government\footnote{Lara Jakes, “As Protests in South America Surged, So Did Russian Trolls on Twitter, U.S. Finds” New York Times, January 19 2020, accessed  December 19 2020, https://www.nytimes.com/2020/01/19/us/politics/south-america-russian-twitter.html}. \\
In this work, we seek to explore the polarization that was observed during the online Twitter discussions throughout the protests. The analysis focuses on two dimensions: Polarization in language and in news consumption patterns. For this purpose, we first separate users by their stance towards the protests and government of each country. We quantify linguistic polarization by training word embeddings on the corpora based on the tweets of users of each stance. We then estimated a translation matrix between both embeddings and explored systematic differences in a translated word and its closest neighbors in the target language. This approach has been recently applied to interpret political polarization in the viewerships of US cable news networks. Our work contrasts with this recent research~\cite{khudabukhsh2020} along the following three main dimensions: (1) different political context (2019 South American Protest instead of US political polarization); (2) different language (Spanish instead of English); and (3) different social media platform (Twitter instead of YouTube). We find a clear polarization in language, mainly manifested along ideological, political or protest-related lines (e.g., Socialism in one group is discussed in similar context to Communism in the other and police in the same way as vandals). We also show that this methodology can be used to mine knowledge, as we find that local political leaders from the protest in one country translate to their counterparts in another  country.\\
In terms of polarization in news consumption, we focus on how confirmation bias manifests in the consumption of news articles based on user stances. The current fractionalized way of consuming news facilitated by social media not only helps to gain a quicker overview current events, but also can lead to confirmation bias \cite{GARRETT2017370}.  For this reason, it is important to explore how it may lead to the manipulation of public opinion by actors with devious motives. By clustering news media based on the homogeneity of their user bases, we find that agencies are clustered geographically and ideologically. Moreover, we find strong evidence of the polarization in users' news consumption patterns and of the presence of ``filter bubbles''. These can have pervasive effects on public discourse and political literacy. \\
This work falls in the emerging field of social cybersecurity which is concerned with the study of disinformation, hate speech and extremism online \cite{carley2020social}. It contributes to the study of polarization in South American politics and linguistic polarization in Spanish during charged political events. 

\section{Literature Review}
\subsection{Polarization in language}
Polarization in political topics has been widely studied in academic literature, focusing especially on partisanship in US politics. The existing research covers aspects like climate change \cite{dana,matt}, gun control/rights \cite{Dora2019Shooting}, Supreme Court confirmations \cite{Darwish2020Kavanaugh}, economic decisions \cite{cy}, congressional votes \cite{keith}, polarization in media \cite{markus}, etc. A recent vein of research has focused in leveraging the knowledge extracted by unsupervised language models to assess polarization. Related work in this area includes: using language models to explore religious polarization in India \cite{shrip}, the study of the presence of human bias in word embedding in social media \cite{aylin,nikhil} and a recent work of word embedding de-biasing \cite{tolga,thomas}. Recently, an approach has been offered to interpret polarization through machine translation \cite{khudabukhsh2020}. This work presented a quantifiable framework with a novel assumption that sub-communities consuming different US cable news networks speak in different \emph{languages}. With this assumption, \cite{khudabukhsh2020} presented a method to quantify similarities (and dissimilarities) between large-scale text data.  As we apply their methodology to assess polarization in language, we defer further explanation of it to Section \ref{sec:pol}. Our use of this framework extends previous results in three key ways: we show that this method is generarizable to (1) different social media platforms (the machine translation-based method has only been applied to YouTube data primarily consisting of user-generated texts on news videos~\cite{khudabukhsh2020, capitol2020}); (2) a language different from English; (3) and a completely different socio-political context of linguistic polarization during South-American protests. 

\subsection{Polarization in news consumption}
A review of recent work that explored the effect of confirmation bias on news consumption, notes that most has centered its cognitive dimension, while the social aspects have hardly been investigated \cite{ling2020confirmation}. There is a need to explore the effects of being exposed to contravening information on in-group interactions, which is the focus of the present work. Most echo chambers encountered empirically are a result of social media practices that exhibit polarized content engagement, rather than exposure \cite{GARRETT2017370}. However, there has been contradictory evidence in this regard, as studies have found limited evidence of challenge avoidance in online settings \cite{dimaggio2003does,flaxman2016filter,gentzkow2011ideological} while consistent evidence of reinforcement seeking. On a similar note, a review of numerous studies on Facebook showed that observed polarization is driven more by selective exposure resulting from confirmation bias than by filter bubbles or echo chambers \cite{spohr2017fake}. 
In the context of the spread of misinformation, polarization has been identified as an important driver for its diffusion. This has been observed in the context of disinformation campaigns around the release of popular Marvel movies \cite{babcock2020pretending,babcock2019diffusion} and in the diffusion of scientific, conspiracy theory and satiric Facebook articles \cite{del2016spreading}. In the later study, confirmation bias was identified as a primary driver for this polarization as it lead to homogeneous clusters of media consumption. This is not specific to Facebook, as the same relationship between polarization and confirmation bias has been observed on YouTube in a similar context \cite{bessi2016users}. In this work, we extend this observation to a different social media platform, namely Twitter.\\


\section{Data Description}
\noindent The dataset was constructed by collecting Twitter data for: Ecuador, Chile, Bolivia and Colombia. More than 500 hashtags and terms were used to collect tweets for the different countries. A special effort was taken to collect conversations around antagonistic positions, by including hashtags that were used by different groups (for and against the different governments). This resulted in over 100 million tweets from 15+ million users. \autoref{t1} presents other descriptive statistics from the data collected for the different countries. 

\begin{table}[!htbp]\centering
\resizebox{0.7\linewidth}{!}{
  \begin{tabular}{l c c } 
\toprule

 &\textbf{ Collection Period (in 2019) }& \textbf{Number of Tweets (Millions)}\\
 \midrule
\textbf{Bolivia}&	October 15 to November 24 	&23.5\\
\textbf{Chile}	&October 10 to November 24 	&59.6\\
\textbf{Colombia}&	November 10 to December 24	&20.4\\
\textbf{Ecuador}&	September 25 to October 24 	&19.1\\

\bottomrule

\end{tabular}
}
\vspace{0.5cm}
\caption{Collection period and number of tweets collected for each country.}\label{t1}
\end{table}
\begin{figure*}
\centering
\includegraphics[width= \textwidth]{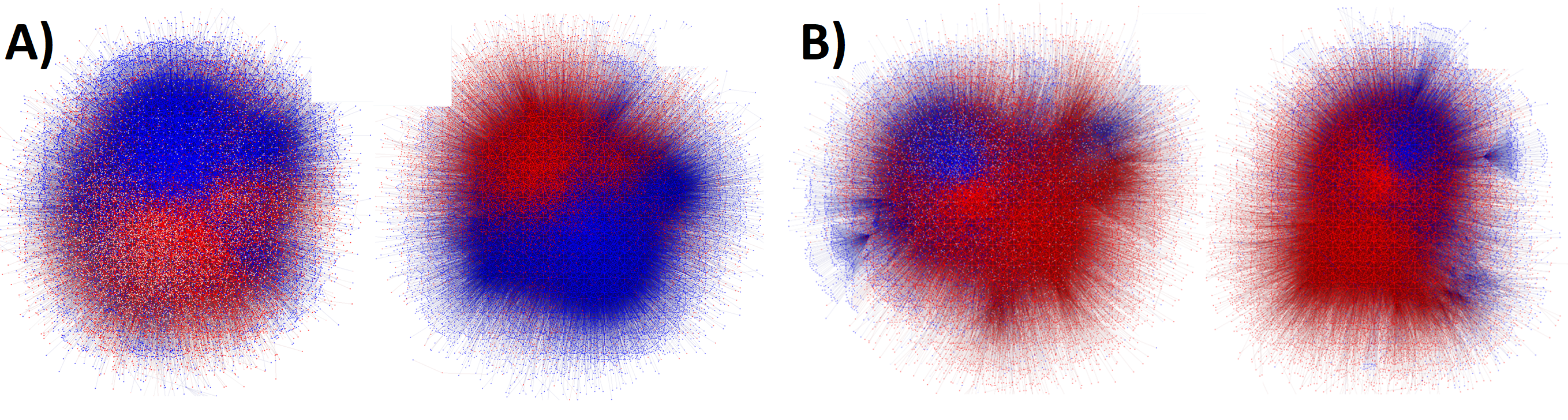}
\caption{User by User response network for A) Bolivia and B) Ecuador. For each panel, the left figure shows the ego-networks of labeled users, including inconsistent users, while the right figure considers consistently labeled users. Users with consistent stances against government are colored red, pro users are colored blue and inconsistent users are colored white.}\label{f1}
\end{figure*}
\vspace{-1ex}

\subsection{Determining user stances}
We constructed a weakly supervised dataset of user stances towards the government and protests that occurred in each country. In order to lend robustness to our dataset construction process, labels were assigned based on two different methods: Hashtag usage and politician retweeting behavior. Our process required determining the stance of a small set of hashtags and politicians. In both cases, the labeling process was carried out by two native Spanish speakers who were familiarized with the events of each protest, based on a sample of their tweets during the protests and in case of disagreements, a consensus label was determined. Labels for each user are assigned based on the probability that they post a tweet with a given stance (based on the defined category). We only considered users that posted at least 10 tweets in our dataset (including retweets).
\subsubsection{Hashtag usage}
 A base set of Hashtags were classified into either pro, anti or neutral to the protests or to the government, based on a sample of the tweets that used them. A hashtag was assigned a stance if it was exclusively used in tweets charged with the corresponding stance. This resulted in 120+ pro and 300 anti-government tags (the remainder being either neutral or not a hashtag). This set was expanded by classifying the most important co-occurring hashtags, following the same methodology. This resulted in 599 pro and 1400 anti-government tags and 1170 pro and 480 anti-protests tags. We found that users that engaged in promoting the protests, sometimes also tweeted things in support of the Venezuelan government that were not directly related to the events of each country. For this reason, the same methodology was used to classify the pro and anti-Venezuela hashtags, resulting in 377 and 108, respectively.\\
 Weak-stance labels were assigned to tweets if they only contain a hashtag from a given stance, if this was not the case the tweet was deemed inconsistent. To assign a stance to a User, we used a probability threshold of 95\%. This implies that at least 95\% of the tweets of a given users were assigned the same stance, either on their tweets or their user descriptions.
 To further validate the stance of the users, stances towards the government and the protests were combined by assigning a consistent stance in favor of the government if a user was assigned a pro stance towards the government and an antagonistic position towards the protests. The opposite was done for the antagonistic stances towards the government. This produced a label distribution that was highly skewed towards anti-government stances.

\subsubsection{Politician retweeting behavior}
We labeled 700 of the major politicians across the countries for the stance towards the government during the protests.  Similar to the methodology described before, users were assigned a stance if they consistently retweeted politicians with the same stance. However, in this case we use a cutoff consistency probability of 90\%. The final user stance was determined by combining the labels obtained by both methodologies, allowing us to validate the stances assigned to the users. There was a 20\% overlap between the labels produced, obtaining inconsistencies that varied from 13-17\% for each of the countries. This provides a measure of the robustness of the labels assigned by our procedure. In total, we were able to assign a weak-stance label users responsible for approximately 20\% of the tweets in the dataset.

\begin{table}[!ht]\centering
  \begin{tabular}{l| c c c c} 
\toprule

 &	\textbf{Bolivia (\%)}	&\textbf{Chile (\%)}	&\textbf{Colombia (\%)}&	\textbf{Ecuador (\%)}\\
 \midrule
\textbf{Consistent Anti-Government}& 42 & 76 & 70 &	66\\
\textbf{Consistent Pro-Government} & 37 & 12 & 22 &	23\\
\textbf{Inconsistent}& 2 & 1 & 3 & 2\\
\textbf{Other}& 19 & 11 & 5 &  8\\

\bottomrule
\end{tabular}
\vspace{0.5cm}
\caption{Distribution of users based on stance towards the government for the collected countries.}\label{t2}
\end{table}

The final distribution of weakly-labeled users and the number of original tweets (not including retweets) posted by them is presented in Table \ref{tbl:users}.
\begin{table}[!ht]\centering
  \begin{tabular}{l c | c c c c} 
\toprule
& & \textbf{Bolivia} & \textbf{Chile} & \textbf{Colombia}&	\textbf{Ecuador}\\
 \midrule
\multirow{2}{*}{\textbf{Consistent Anti-Government}} & Users & 56\,146 & 208\,632 & 85\,923 &	65\,079\\
& Tweets & 1\,483\,202 & 5\,671\,474 & 1\,598\,679 & 954\,508\\
  \cmidrule{2-6}
\multirow{2}{*}{\textbf{Consistent Pro-Government}} & Users & 48\,794 & 32\,052 & 26\,455 &	23\,463\\
& Tweets & 755\,734 & 2\,165\,908 & 927\,769 & 432\,544\\
\bottomrule
\end{tabular}
\vspace{0.5cm}
\caption{Number of Weakly-Labeled Users and their original tweets (not including retweets)}\label{tbl:users}
\end{table}

\subsection{Exploring community interaction}
\begin{figure}
\centering
\includegraphics[width=0.85\columnwidth]{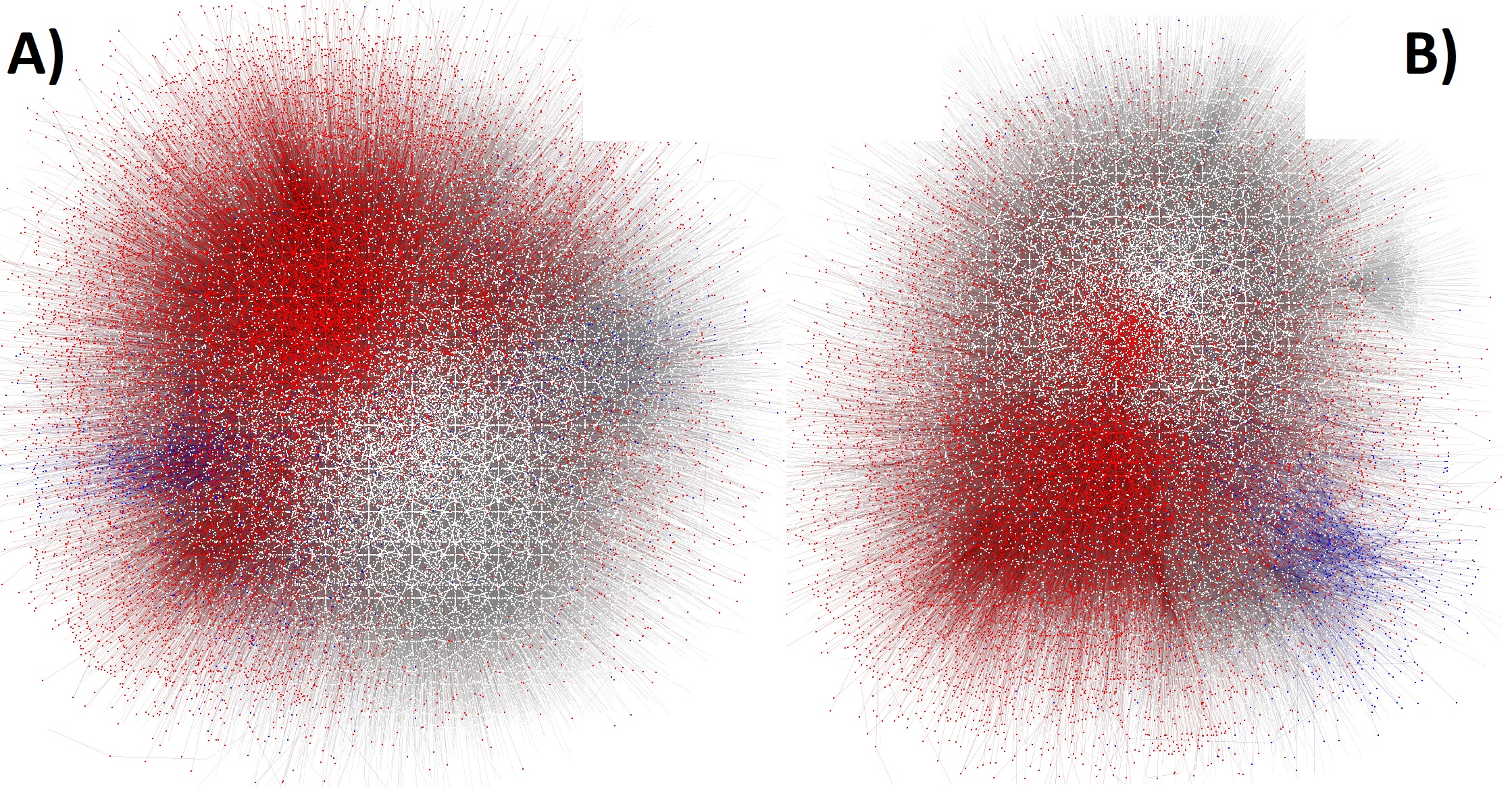}
\caption{User by User response network of consistently labeled users for A) Bolivia and B) Ecuador. Nodes are colored red if they support the Venezuelan government, blue if they were against and white if no label is assigned based on the methodology described. }\label{f2}
\end{figure}

We explored community interaction between pro and anti-government movement through the user by user response network. Two users are linked in this network if they replied each other either by direct responses or quotes. \autoref{f1} presents the ego-network of the labeled users (including unlabeled users) and the subgraph that only includes users that were labeled consistently for the country of Bolivia and Ecuador. As observed, among the consistently labeled users, most communications occur between users of the same stance and the network clearly clusters based on the user’s stance towards the government. However, when we consider unlabeled and inconsistently labeled users, we see that the network becomes more integrated, with unlabeled users serving as bridges between the communities. Similar effects are observed with the other countries.\\
Further, we explore the stance of the users defined before toward the Venezuelan government. As shown in \autoref{f2}, a significant percentage of users that favored the Bolivian government (ideological ally of the Venezuelan regime) also favored the Venezuelan government. However, a much smaller share of the users that were against the government Bolivian government engaged in tweets that revealed their stance towards the Venezuelan government. A similar pattern is observed for Ecuador (and all other countries not shown in the figure). This suggest that there is a more active presence of Venezuelan accounts promoting the Bolivian government, or the protests in the case of Ecuador and other countries studied, an observation that is corroborated by their news consumption patterns, discussed below. 

\section{Linguistic Polarization}\label{sec:pol}

We explored linguistic polarization among both communities by applying the methodology developed in \cite{khudabukhsh2020}. Their work was applied to English YouTube comments in the context of polarization in US cable news networks viewerships, but as we show in this work, the methodology is more generalizable and robust to differences in colloquialisms. In~\cite{khudabukhsh2020}, the authors presented a machine translation based framework to quantify the differences between large scale social-media discussion data sets. This framework assumes that two sub-communities (e.g., the sub-community favoring the protest and the sub-community opposing the protest) are speaking in two different \emph{languages} (say, $\mathcal{L}_\mathit{pro}$ and $\mathcal{L}_\mathit{against}$) and obtains single-word translations using a well-known machine translation algorithm~\cite{SmithTHH17}. Since $\mathcal{L}_\textit{pro}$ and $\mathcal{L}_\textit{against}$ are both in fact Spanish, ideally, any word $w$ in $\mathcal{L}_\mathit{pro}$ should translate to itself in $\mathcal{L}_\mathit{against}$. However, when a word $w_1$ in one language translates to a different word $w_2$ in another, it indicates $w_1$ and $w_2$ are used in similar contexts across these two \emph{languages} signalling (possible) disagreement. These disagreed pairs present a quantifiable measure to compute differences between large scale corpora as greater the number of disagreed pairs the farther two sub-communities are. 

We now present a formal description of this framework. Let our goal be to compute the similarity between two languages, $\mathcal{L}_\mathit{source}$ and $\mathcal{L}_\mathit{target}$, with vocabularies $\mathcal{V}_\mathit{source}$ and $\mathcal{V}_\mathit{target}$, respectively. Let $\emph{translate}(w)^{\mathcal{L}_\mathit{source} \rightarrow \mathcal{L}_\mathit{target}}$ denote a single word translation of $w \in \mathcal{V}_\mathit{source}$ from $\mathcal{L}_\mathit{source}$ to $\mathcal{L}_\mathit{target}$. The similarity measure between two languages along a given translation direction computes the fraction of words in $\mathcal{V}_\mathit{source}$  that translates to itself, i.e.,\\
\large{
\emph{Similarity}\hspace{0.03cm}($\mathcal{L_\mathit{source}}, \mathcal{L_\mathit{target}})$ = $\frac{\Sigma_{w \in \mathcal{V}_\mathit{source}} \mathbbm{I}(\emph{translate}(w)^{\mathcal{L}_\mathit{source} \rightarrow \mathcal{L}_\mathit{target}} = w)}{|\mathcal{V}_\mathit{source}|}$
}.
\normalsize
The indicator function returns 1 if the word translates to itself and 0 otherwise. The larger the value of \emph{Similarity} ($\mathcal{L_\mathit{source}}, \mathcal{L_\mathit{target}})$, the greater is the similarity between a language pair.


We constructed $\mathcal{L}_\textit{pro}$ and $\mathcal{L}_\textit{against}$ by combining all the main tweets by a user of a given stance (this includes tweets not related to the protests). We made sure to include a retweeted tweet only once (instead of the millions of times it might have been retweeted). For each of constructed corpora, we trained a 100-dimensional FastText embedding \cite{bojanowski2017enriching}, with basic pre-processing that included removing URLs, mentions and punctuation.  
Next, we  translate the top 10k words in one language to the other and examine disparities. The translation scheme learns a transformation matrix $T$ between the two embeddings, such that the product with the embedding of the source produces the target. We followed the same experimental protocols presented in~\cite{khudabukhsh2020}: (1) we used stop-words as anchors for the translation as these are the most likely to maintain their meaning across the different groups; and (2) given the imbalanced size of the two corpora, we sub-sampled the majority community (against the government) to match the size of the smallest community across the countries (Ecuador's Pro-Government community). This guaranties a fair comparison between the two communities. 

Due to space constraints, in Table \ref{tbl:trans} we provide several notable examples of the polarized language observed between the two communities only for Bolivia and Ecuador (the results are consistent across all the countries studied). We find a polarization in language, mainly manifested along ideological, political and protest-related lines. Terms related to left-leaning ideologies in one community tend to be discussed in similar contexts as right-leaning terms (e.g. Socialism translates to Communism); terms related to the protests and protesters are discussed in similar context as terms related to law and order (an informal term for police translates to vandal). Importantly, we find that the motivations for the protests translate to each other. For example, in the case of Ecuador, 883\footnote{This refers to the decree 883 which proposed austerity measures and started the protests in the country} translates to derogate which was one of the calls of the protests movements (a similar pattern is shown for Bolivia with the "overthrow" term). In the case of Bolivia, overthrowing (accusation made by Evo Morales against the protesters) translates to defeat (which refers to the electoral victory the opposition claimed to have achieved). Finally, opposition leaders are discussed in similar context as government representatives. 

\begin{table}[h!]
\small
\centering
  \begin{tabular}{l c c c}
\toprule
\multirow{2}{*}{\textbf{Country}} & \multirow{2}{*}{\textbf{Theme}} &	$L_{pro}$ & $L_{against}$\\
 & & (translation) & (translation) \\
 \midrule
\multirow{18}{*}{Ecuador} & \multirow{6}{*}{Idiological} & Comunismo   & Neoliberalismo \\
          & & (comunism) & (neoliberalism) \\
          & & Socialismo & Capitalismo \\
          & & (socialism) & (capitalism)\\
          & & Populista  & Autoritario \\ 
          & & (populist) & (authoritarian) \\
  \cmidrule{2-4}
          & \multirow{6}{*}{Protest} &  Chapas   & Vandalos \\
          & & (police -informal-) & (vandals) \\
          & & Protestantes & Policias \\
          & & (protesters) & (police)\\
          & & 883  & Derogar \\ 
          & & (name of policy) & (derogate) \\  
  \cmidrule{2-4} 
          & \multirow{6}{*}{Political} &  Indigena & Insurgencia \\
          & & (Indigenous) & (insurgency) \\
          & & Morenista & Correista \\
          & & (president suporter) & (supporter of Correa)\\
          & & Borrego  & Correista\\
          & & (sheep) & (supporter of Correa)\\
 \midrule
 \midrule
\multirow{18}{*}{Bolivia} & \multirow{6}{*}{Idiological} & Comunismo   & Facismo \\
          & & (communism) & (fascism) \\
          & & Socialista & Capitalista \\
          & & (socialist) & (capitalist)\\
          & & Populista  & Fascista \\ 
          & & (populist) & (facist) \\
  \cmidrule{2-4}
          & \multirow{6}{*}{Protest} &  Golpista   & Fascista \\
          & & (coup plotter) & (fascist) \\
          & & Derrocar & Derrotar \\
          & & (overthrow) & (defeat)\\
          & & Molotov  & Lacrimogena \\ 
          & & (molotov) & (tear gas) \\  
  \cmidrule{2-4} 
          & \multirow{6}{*}{Political} &  evomorales & Criminal \\
          & & (ex-president) & (criminal) \\
          & & Masista & Maleante \\
          & & (opposition supporter) & (malefactor)\\
          & & Masistas  & Golpistas\\
          & & (opposition supporters) & (coup plotter)\\ 
\bottomrule
\end{tabular}

\caption{Notable instances of linguistic polarization by topic for Ecuador and Bolivia.}
\label{tbl:trans}
\end{table}

\begin{table}[ht]
\centering
\begin{tabular}{c c| c| c| c| c|}

&\multicolumn{1}{c}{} &\multicolumn{4}{c}{$L_\textit{target}$ ($L_\textit{against}$)}\\
&\multicolumn{1}{c}{} & \multicolumn{1}{c}{Bolivia} & \multicolumn{1}{c}{Chile}&\multicolumn{1}{c}{Colombia}& \multicolumn{1}{c}{Ecuador}\\\cline{3-6}

\multirow{4}{*}{$L_\textit{source}$ ($L_\textit{pro}$)}&Bolivia  & 49.9 (0.6)&-&-&- \\ \cline{3-6}
  &Chile & - &63.1 (0.6)&-&-\\ \cline{3-6}
 &Colombia&- &-&57.1 (0.6)&-\\ \cline{3-6}
 &Ecuador&-&-&-&56.0 (1.3)\\ \cline{3-6}

\end{tabular}
\vspace*{0.5em}

\begin{tabular}{c c| c| c| c| c|}

&\multicolumn{1}{c}{} &\multicolumn{4}{c}{$L_\textit{target}$ ($L_\textit{pro}$)}\\
&\multicolumn{1}{c}{} & \multicolumn{1}{c}{Bolivia} & \multicolumn{1}{c}{Chile}&\multicolumn{1}{c}{Colombia}& \multicolumn{1}{c}{Ecuador}\\\cline{3-6}

\multirow{4}{*}{$L_\textit{source}$ ($L_\textit{pro}$)} &Bolivia  & - &40.1 (0.9) & 37.0 (0.6) & 38.6 (0.7)\\ \cline{3-6}
  &Chile & 40.1 (0.9)& - & 58.7 (0.9) & 52.6 (1.4) \\ \cline{3-6}
 &Colombia& 37.0 (0.6) & 58.7 (0.9)  & - & 49.9 (0.9)\\ \cline{3-6}
 &Ecuador& 38.6 (0.7) & 52.6 (1.4) & 49.9 (0.9)&-\\ \cline{3-6}

\end{tabular}

\vspace*{0.5em}

\begin{tabular}{c c| c| c| c| c|}

&\multicolumn{1}{c}{} &\multicolumn{4}{c}{$L_\textit{target}$ ($L_\textit{against}$)}\\
&\multicolumn{1}{c}{} & \multicolumn{1}{c}{Bolivia} & \multicolumn{1}{c}{Chile}&\multicolumn{1}{c}{Colombia}& \multicolumn{1}{c}{Ecuador}\\\cline{3-6}

\multirow{4}{*}{$L_\textit{source}$ ($L_\textit{against}$)} &Bolivia  & - & 46.1 (0.8) & 43.4 (1.2) & 47.2 (0.6) \\ \cline{3-6}
  &Chile & 46.1 (0.8) & - & 58.7 (0.9) & 52.6 (1.4) \\ \cline{3-6}
 &Colombia& 43.4 (1.2) & 58.7 (0.9) & - & 49.9 (0.9)\\ \cline{3-6}
 &Ecuador& 47.2 (0.6) & 52.6 (1.4) & 49.9 (0.9) & -\\ \cline{3-6}
\vspace*{0.5em}
\end{tabular}

\caption{Pairwise similarity (in percentage points) between languages computed for: Top) pro and against protest communities in different countries, Middle) pro-protest communities in different countries, and Bottom) against-protest communities in different countries. All similarity metrics are constructed with corpus of approximately the same size. Standard deviations are presented in parenthesis, most statistics are more than 2 deviations away from each other. The evaluation set, is computed by concatenating the corpora and taking the top 5K words ranked by frequency.}\label{t-comparison}
\end{table}
\pagebreak
The mean and standard deviation of the pairwise similarity between languages in pro and against protest communities within countries are shown in the top section of Table \ref{t-comparison}. The sub-sampling process was repeated 6 different times, given the randomness of this process.  As we can observe, the largest linguistic polarization was observed in Bolivia, this difference being significant at a 95\% threshold. Moreover, to test the robustness of the methodology to differences in dialects, we apply it to compare stances between the countries. The middle and bottom sections of Table \ref{t-comparison} include the pairwise similarity of languages for pro- and against-government communities between countries. First, notice that the similarities between the countries are significantly lower in the pro-government case, as this community is considerably smaller than its counterpart (which is expected because corpus size is one of the most important contributing factors to ensure the quality of word embedding). However, even with this caveat in mind, the language similarity is lower for pairs that contrast Bolivia to other countries. This is consistent with the fact that Bolivia is the only country which has a left-leaning government and hence right-leaning protesters. Moreover, even with the added noise of local colloquialisms, we are still able to recover the ideological polarization between protest movements, when compared to Bolivia  (``neoliberalism'' still translates to ``communism'' or ``socialist'' to ``nazi''). These differences do not exist when comparing protests movements with similar ideological motivations (e.g. ``socialist'' in Ecuador translates to ``socialist'' in Chile). Importantly, we can use this methodology to mine knowledge from the corpora, as we find that local political leaders from the protest in one country translate to their counterparts in the other country (e.g. ``Correa'' in Ecuador translates to ``Petro'' in Colombia -- an opposition leader).

\section{Polarization in News Consumption}
We first constructed a data set of the regional and local news agencies operating in each country. For the purpose of exploring regional influence campaigns on the protests, we included agencies from Venezuela and Russia that predominately operate in the region. The number of news agencies from each country is shown below in \autoref{t1}. For each of them we collected, their name, Twitter handle and the URL for their main domain. However, news articles identified in our data set cover topics ranging from the protests to sports. When studying the polarization of news consumption the political event, it was important to first filter out the tweets which are irrelevant to the protests.  It is not obvious if a tweet from news media is relevant to the ongoing protests in each country or not, but many tweets in our data set contain the URL of an article that they reference. For this reason, we determined the relevance to the protest of a small set of the 900 most tweeted URLs in our dataset for any of the countries studied. We complemented this dataset with an additional set of URLs labeled by extracting subsection metadata from them. If the subsection referenced sports, culture or technology, the URL was labeled as irrelevant to the protests. Then, we assigned the URL label to any tweet that used it. The final sample distribution are presented in Table \ref{t:sample}. We note that even though we are able to assign a label to more than 100k tweets, most of them contained duplicated text (as news media tend to tweet the same thing multiple times). The classification was done with the unduplicated dataset.

\begin{table}[!ht]
\centering
  \begin{tabular}{l c} 
\toprule

 \textbf{Country} &	\textbf{\# Agencies}\\
 \midrule
Bolivia &52\\
Chile & 95\\
Colombia& 69\\
Ecuador &99\\
Regional* & 10\\
Russia & 2\\
Venezuela & 57\\

\bottomrule

\end{tabular}
\vspace{0.5cm}
\caption{Number of news agencies in each country. *The regional category includes regional Venezuelan and Russian media among others.}\label{t1}
\end{table}

\noindent We built a CNN text classifier \cite{kim2014convolutional} using 300 dimensional FastText embeddings trained on the combined datasets (both by stance and country) used to analyze the language polarization. We used 100 filters on 3 layers with filter sizes 3, 4 and 5 and a dropout rate of 50\%.  We achieved an accuracy and F1-score of 92\% in a held-out test set.

\begin{table}[!ht]
\centering
\begin{tabular}{l c c} 
\toprule
 \textbf{Category} &	\textbf{\# of Tweets} & \textbf{\# of Tweets (Unduplicated)}\\
 \midrule
Relevant & 78,318 & 7,177\\
Not Relevant & 53,671 & 8,156\\

\bottomrule
\end{tabular}
\vspace{0.5cm}
\caption{Distribution of the labeled tweets}\label{t:sample}
\end{table}

After predicting the labels of tweets (relevant or irrelavant to the protests), we obtain a dataset of 1,024,166 relevant and 675,496 irrelevant tweets. The distribution of the data set is shown in \autoref{t3}. For our analysis below, we filter out the irrelevant tweets and focus only on the tweets that are relevant to the protests. \\
\begin{table}[!ht]
\centering
\begin{tabular}{l c c} 
\toprule
 \textbf{Category} &	\textbf{\# of Tweets} & \textbf{\# of Users}\\
 \midrule
 
Relevant &1,024,166& 276,754\\
Not Relevant &675,496 & 247,324\\

\bottomrule
\end{tabular}
\vspace{0.5cm}
\caption{Distribution of the tweets with predicted labels and users}\label{t3}
\end{table}

\begin{figure}[!ht]
\centering
\includegraphics[width= 0.5\columnwidth]{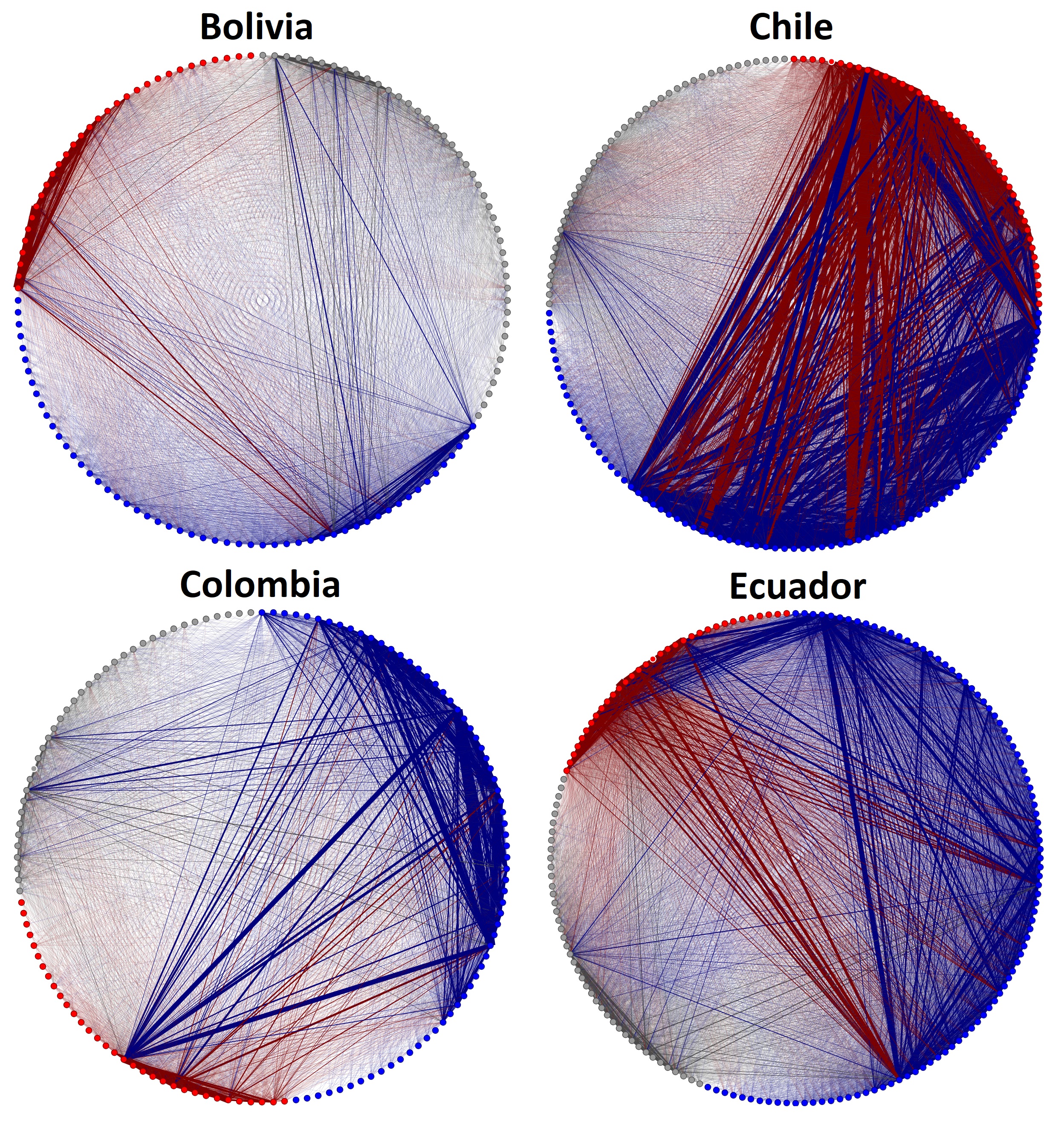}
\caption{The cluster of the news agencies. The colors indicate different communities, where a) blue are local media; b) red are regional Russian and Venezuelan Media; and c) green colors are local Venezuelan media. Edges between dots represent common users are shared by the news media on the two ends. }\label{f3}
\end{figure}

\paragraph{\textbf{Community detection among News Agencies}}
After filtering out the irrelevant tweets, we explore the community of news agencies based on the homogeneity of their user bases. 
We apply Louvain community detection to cluster the news agencies. Formally, a bipartite graph is a triple $G = (A, B, E) $ where $A = \{a_i | i = 1, 2, ..., n_A\}$ and $B = \{b_j | j = 1, 2, ..., n_B\}$ are two disjoint sets of vertices, and $E$ is a set of edges where they only exist between set $A$ and $B$. The bipartite graph $G$ could be described as a matrix $M$ for which
\begin{align*}
    M_{i,j} = \begin{cases}
    1 , & \text{if there is an edge between } a_i\text{ and } b_j\\
    0,              & \text{otherwise.}
\end{cases}
\end{align*}
As shown in \autoref{f3}, media are divided into three clusters: a) major local media (blue), b) regional Russian and Venezuelan media (red), including local left-leaning media, and c) local Venezuela media (gray). Note that agencies are not only clustered geographically, but also ideologically as the local media cluster includes the major media organizations in each country, which, as mentioned before, were reportedly more likely to support the local governments (with the exception of Bolivia where the situation is reversed). It is noteworthy that there exists a cluster of local Venezuelan media, which confirms the observation presented before that there was an important presence of Venezuelan accounts active during the protests in each of the countries. The clustering results confirms our hypothesis that media with similar stances are more likely to share homogeneous users' base providing evidence for of ``filter bubbles'' in users' news consumption behavior. In what remains of the analysis, we do not consider the local Venezuelan cluster. 

\subsection{Quantifying the polarization}
\begin{figure}[!htbp]
  \centering
  \subfigure{\includegraphics[scale=0.19]{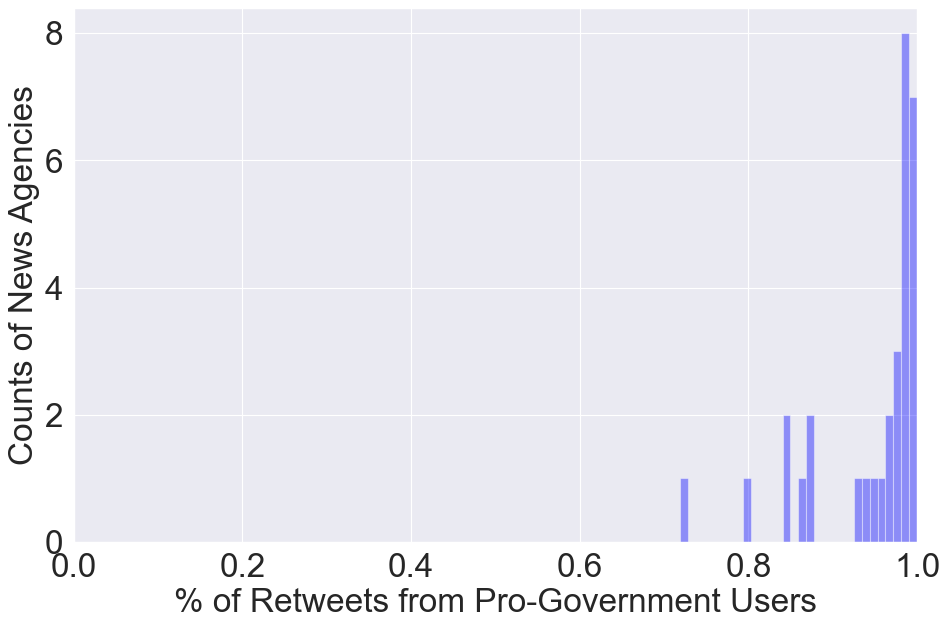}}\quad
  \subfigure{\includegraphics[scale=0.19]{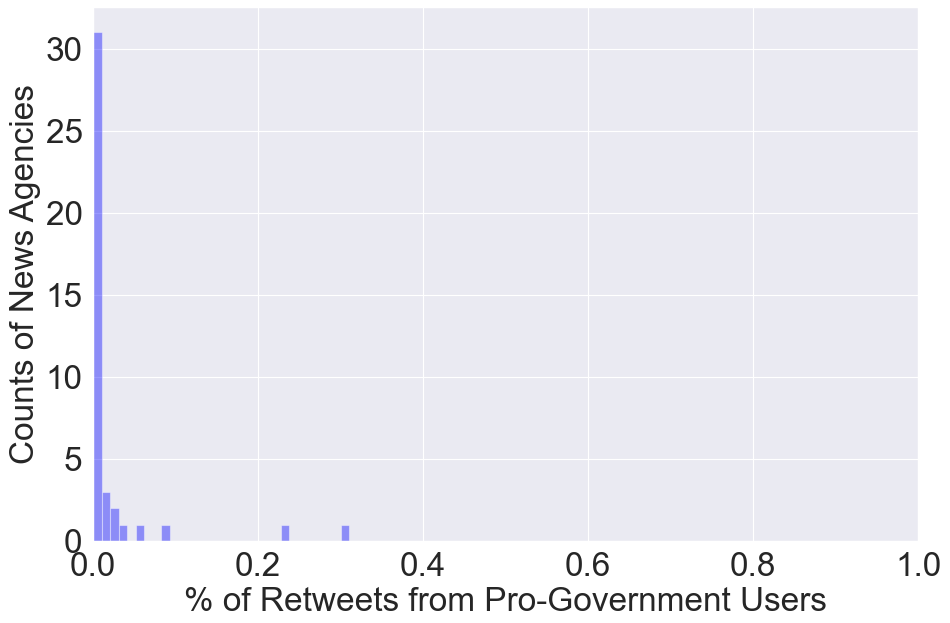}}\quad
  \subfigure{\includegraphics[scale=0.19]{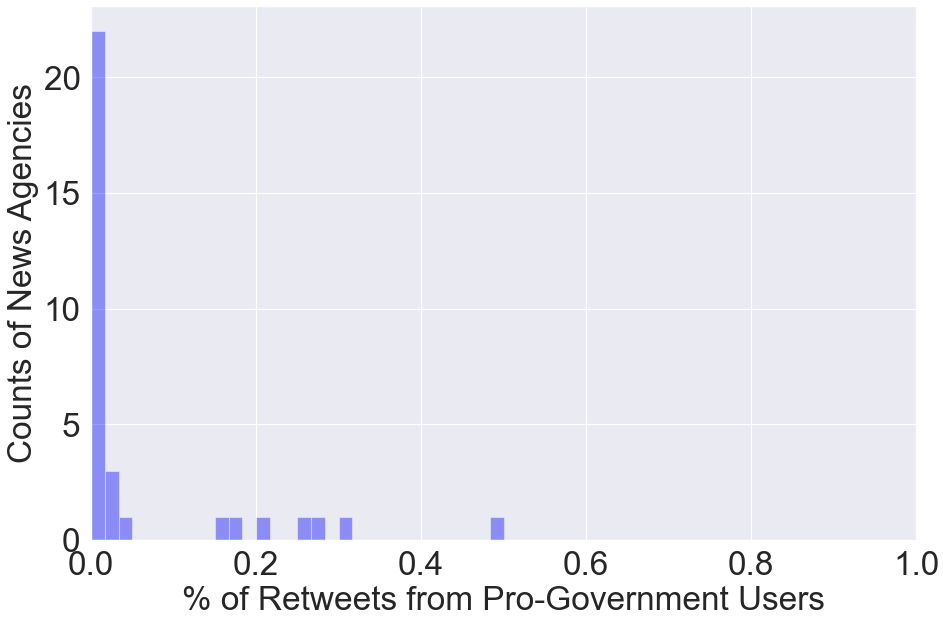}}\quad
  \subfigure{\includegraphics[scale=0.19]{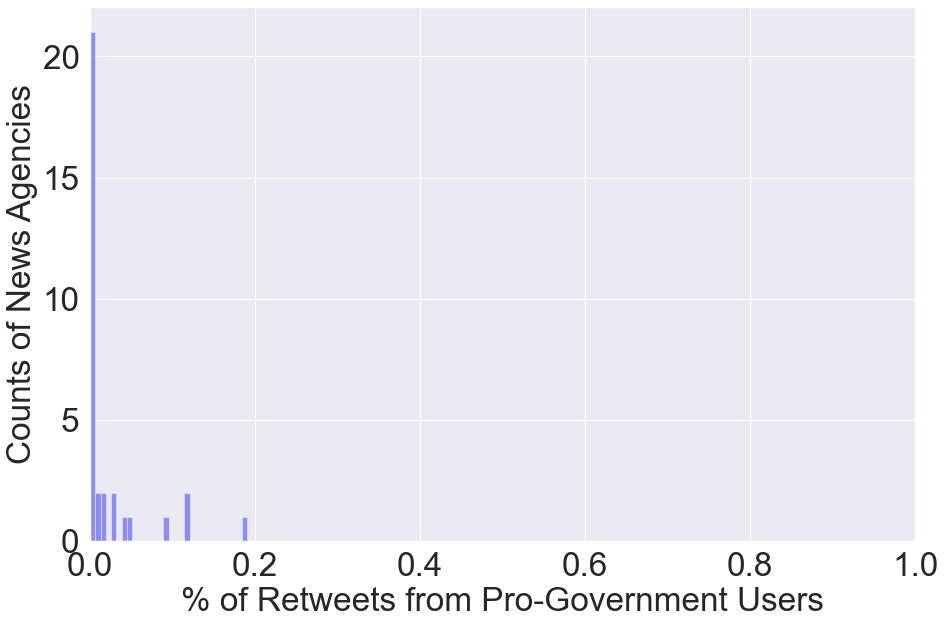}}
  \caption{Histograms of ratio of retweets from pro-government users for regional Russian and Venezuelan media: for country Bolivia (up left), Chile (up right), Colombia (bottom left) and Ecuador (bottom right)}\label{fb}
\end{figure}

\begin{figure}[!htbp]
  \centering
  \subfigure{\includegraphics[scale=0.19]{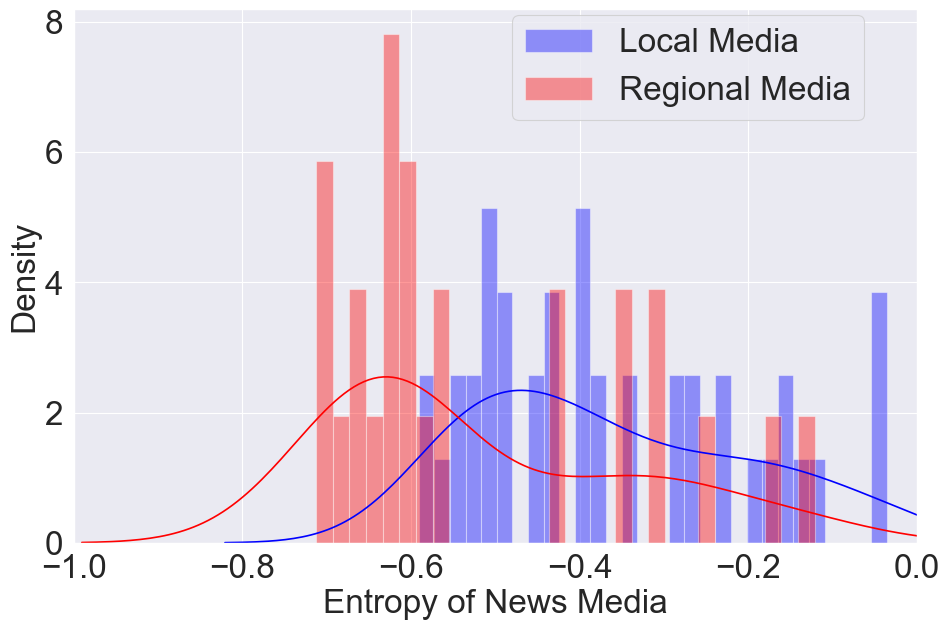}}\quad
  \subfigure{\includegraphics[scale=0.19]{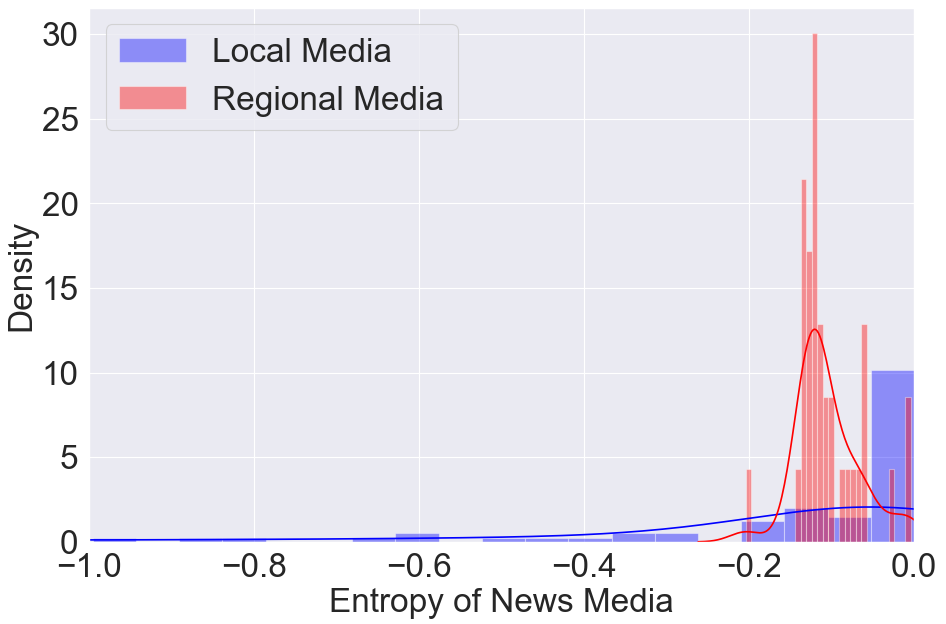}}\quad
  \subfigure{\includegraphics[scale=0.19]{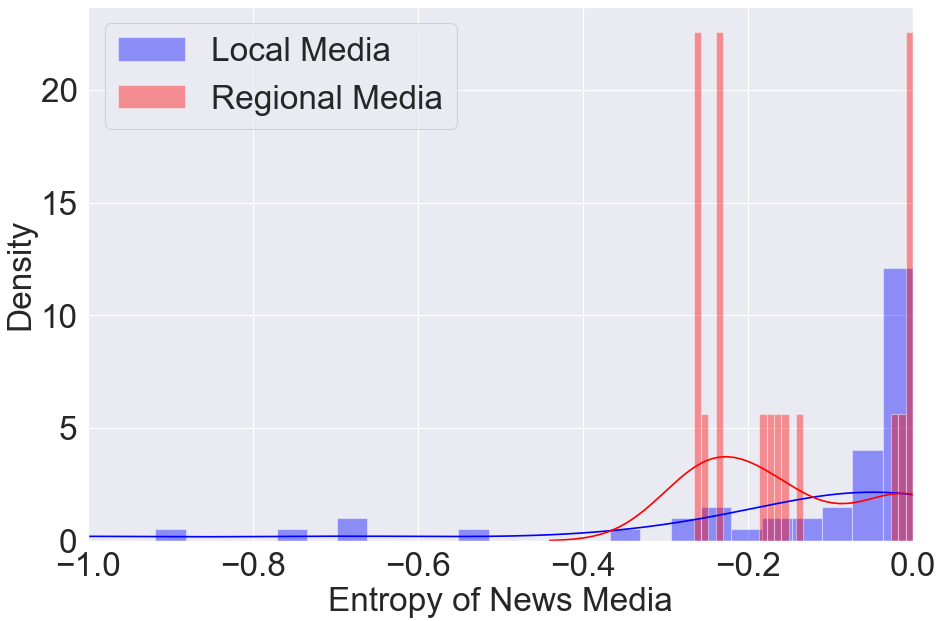}}\quad
  \subfigure{\includegraphics[scale=0.19]{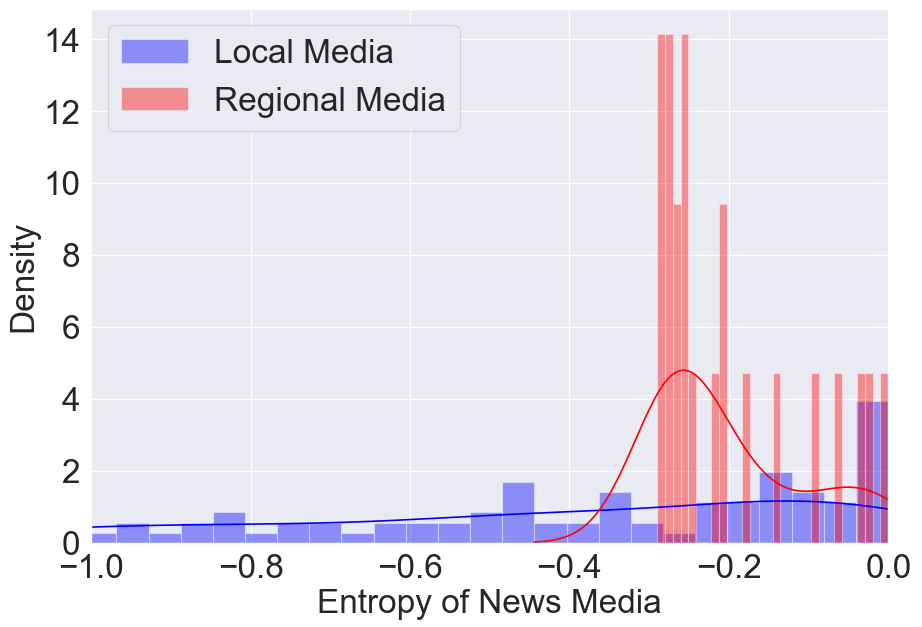}}
  \caption{Distribution of relative entropy for local (blue cluster) and regional Russian and Venezuelan media (red cluster) in Bolivia (up left), Chile (up right), Colombia (bottom left) and Ecuador (bottom right).}\label{n}
\end{figure}

We next measure the the diversity in the user bases of the news agencies in the region to evaluate the polarization in news consumption by users with different stances. \autoref{fb} shows the distribution of the ratio of retweets from pro-government users in Bolivia, Chile, Colombia and Ecuador for the regional Russian and Venezuelan news media. For regional left-leaning media, most of the retweets come from anti-government users in Chile, Colombia and Ecuador. Whereas in Bolivia, most retweets of these news agencies come from pro-government users. This is consistent with their political orientation and their documented support of the Bolivian government \cite{savageRT}, while Ecuador, Chile and Colombia have more right-leaning governments.  The results not only provide evidence of polarization, but also for the ``filter bubble" hypothesis of information diffusion, where users are more likely to retweet news aligning with their stances.\\
We further explore if the observed level of polarization described before is only accounted by the asymmetries in the user stances observed in each country. For this purpose, we compute the relative entropy for each news agency in the clusters of local and regional Russian and Venezuelan media. For each news agency $n$, its relative entropy is defined as follows:
\begin{equation}
H(n) = - p_{pro}*log\frac{p_{pro}}{g_{pro}} - (1-p_{pro})*log\frac{1-p_{pro}}{1-g_{pro}}
\end{equation}
where $p_{pro}$ is the ratio of retweets for news media $n$ from pro-government users and $g_{pro}$ is the overall ratio of pro-government users in that country. The relative entropy $H(n)$ evaluates how $p_{pro}$ differs from $g_{pro}$ - the lower the value of $H(n)$, the more disproportional the level of polarization is with respect to the asymmetries observed in the stance distribution. The maximum possible value 0 of $H(n)$ is obtain when a news organization's user base matches the the stance distribution for the country. \autoref{n} shows the distribution of relative entropy for news media in the clusters of local media and regional Russian, Venezuelan media for Bolivia, Chile, Colombia and Ecuador. We observe that for all countries, regional media from Russia and Venezuela are disproportionately polarized. Moreover, in the case of Bolivia, we observe the highest level of relative polarization, an observation that is consistent with the language polarization levels described in the previous section.
%

\subsection{Assessing confirmation bias through news media transitions}
We finally address the way users navigate among the different clusters of news media. We adapt the methodology developed in Koutra et al. 2015\cite{koutra2015}, where they examined the transition behavior of users when browsing websites related to the topic of gun rights/control. Specifically, we seek insights about how the political orientation of news media influences the type of news agencies that users browse or retweet next, in the context of a polarising topic like the protests. We explore the type of news media tweets or articles that users are more likely to retweet after consuming news supporting or objecting the protest. For each user, we represent her retweeting history as a Markov chain of the cluster of agencies the retweeted news come from. Then, we describe the distribution of the transition probabilities by an n-state transition matrix $P_n$, with elements $p_{ij} = \text{Prob}(X_{t+1} = j|X_t = i)$. We note that the row-wise sums are equal to 1. Specifically, state 0 represents when the user retweets news from local news media, state 1 from regional Russian and Venezuelan media. The transition matrix for Bolivia, Chile, Colombia and Ecuador are shown in \autoref{tb}. To make sense of the underlying trends of this matrix, we employ mobility indices that have been widely used in economics and sociology.

\begin{table}
[!ht]\centering

 \begin{tabular}{l l c c } 
\toprule
 Country &&Local media &	Regional media\\
\midrule

\multirow{2}{*}{Bolivia} &Local media&94.15\% & 5.85\%\\
&Regional media&2.79\% & 97.21\%\\
\midrule
\multirow{2}{*}{Chile} &Local media&66.68\% & 33.32\%\\
&Regional media&21.2\% & 78.8\%\\
\midrule
\multirow{2}{*}{Colombia} &
Local media&91.75\% & 8.25\%\\
	&Regional media&17.96\% &82.04\%\\
\midrule
\multirow{2}{*}{Ecuador} &Local media &83.27\% &16.73\%\\
&Regional media& 7.85 \% &92.15\%\\
\bottomrule
\end{tabular}
\caption{Transition matrix for Bolivia, Chile, Colombia and Ecuador}\label{tb}
\end{table}

\noindent We employ the Summary Mobility Indices, which describe the
direction of the mobility:
\begin{itemize}
    \item Immobility Ratio: IR =$\sum_{i=1}^n p_{ii}/n$
    \item Moving Up (Left): MU (ML) =
$\sum_{i<j} p_{ij}/n$
\item Moving Down (Right): MD (MR) =
$\sum_{i>j} p_{ij}/n$
\end{itemize}
where $n$ is the number of states. As mentioned previously, the cluster of Russian and Venezuelan media is left-leaning media. We therefore call the index of moving up (from local media to Russian, Venezuelan media) as moving left (politically), and the index of moving down as moving right (politically). 

\begin{table}
[!htbp]\centering

  \begin{tabular}{l l l l} 
\toprule
Country& IR&	ML& MR\\
 \midrule
Bolivia &95.68\% &2.93\%&1.40\%\\
Chile& 72.74\% &16.66\%&10.6\%\\
Colombia&86.90\%&4.13\%&8.98\%\\
Ecuador &87.72\%&8.37\%&3.93\%\\

\bottomrule

\end{tabular}

\caption{Summary Mobility Indices}\label{sum}
\end{table}
The Summary Mobility Indices for different countries are included in \autoref{sum}. We note that the majority transitions are to the same state, which indicates users tend to stay in the community (bubble) where the stances of the information align with their own.  The likelihood of transitioning out of a user's community (bubble) is generally low. In this regard, the transition matrices provide evidence for the presence of ``filter bubbles'' in the midst of the protests event and confirmation bias in the way users choose to share content. Moreover, the observed polarization levels in news transitions are consistent with what was described for the language polarization in the last section; with Chile showing the lowest level of polarization and Bolivia the highest, and Colombia and Ecuador showing similar levels. In addition, we note that, with the exception of Colombia, higher percentage of transitions occur in the direction moving to the left (politically), with ML $>$ MR, which suggests in the three countries it would be more difficult for users with left stances to consume news on their right (politically) than the other way around. 

\section{Discussion}

In this work we explored polarization in user behavior, based on their stance towards the government during the 2019 South American protests. We focused on polarization in language and in news media consumption. We find that user discussion, based on responses, clearly clusters according to the stance of the users. Moreover, there is an important presence of accounts that promote the Venezuelan government and are active in the discussions supporting the protests (or in the case of Bolivia, supports the government). In terms of the language used by each of the communities, we find a clear polarization, mainly manifested along ideological, political and protest-related lines. This pattern is maintained across all countries. We also show that we can mine knowledge from the corpora, as local political leaders from the protest in one country translate to their counterparts in the other country. 
In addition, we find strong evidence of polarization in news consumption and information diffusion by users, consistent with their stances towards the government. The news media in our data set clearly clusters with the political stances of their content. We find consistent evidence of filter bubbles in news consumption on Twitter, as users tend to stay in the community of news media that shares information they are more like to agree with. But there is a tendency for users to transit into consuming left-leaning news. Moreover, we show the important role that regional Russian and Venezuela news outlets like RT en Español and TeleSUR, played in the social media discussion of the protests throughout the region. This shows how effective these outlets have been in gathering an audience of left-leaning users in the region, an initiative that has been identified by other studies of these networks \cite{savageRT}. Finally, we see that along both dimensions of polarization explored, we find consistent results, with Chile showing the lowest level of polarization and Bolivia the highest, and Colombia and Ecuador showing similar levels.

\section*{Acknowledgments}
This work was supported in part by the Knight Foundation and the Office of Naval Research grant N000141712675. Additional support was provided by the Center for Computational Analysis of Social and Organizational Systems (CASOS) and the Center for Informed Democracy and Social Cybersecurity (IDeaS). Ramon Villa-Cox is sponsored in part by Secretaría de Educación Superior, Ciencia, Tecnología e Innovación (SENESCYT), Ecuador. The views and conclusions contained in this document are those of the authors and should not be interpreted as representing the official policies, either expressed or implied, of SENESCYT, the Knight Foundation, Office of Naval Research or the U.S. government.

\bibliographystyle{unsrt} 

\end{document}